\documentstyle[12pt]{article}
\begin{document}
\begin{titlepage}
\rightline{SNUTP/96-72}
\def\today{\ifcase\month\or
        January\or February\or March\or April\or May\or June\or
        July\or August\or September\or October\or November\or December\fi,
  \number\year}
%\rightline{\today}
\rightline{hep-th/9701129, revised version, March}
\vskip 1cm
\centerline{\Large \bf Compactifications of F-Theory }
\centerline{\Large \bf on Calabi-Yau Threefolds at Constant Coupling}
\vskip 2cm
\centerline{\sc Changhyun Ahn \footnote{chahn@nms.kyunghee.ac.kr}
 and Soonkeon Nam \footnote{
nam@nms.kyunghee.ac.kr}}
\vskip 1cm
\centerline {{\it Department of Physics and}}
\centerline{{\it Research Institute for Basic Sciences,}} 
\centerline {{\it Kyung Hee University,}}
\centerline {{\it Seoul 130-701, Korea}}
\vskip 1cm
\centerline{\sc Abstract}
\vskip 0.2in
Generalizing the work of Sen, we analyze special points in the
moduli space of the compactification of the F-theory on elliptically
fibered Calabi-Yau threefolds where the coupling remains constant.
These contain points where they can be realized as orbifolds of six torus
${\bf T}^6$ by ${\bf Z}_m \times {\bf Z}_n (m, n=2, 3, 4, 6)$.
At various types of intersection points of singularities,
we find that the enhancement of gauge symmetries arises from the intersection
of two kinds of singularities.
We also argue that when we take the Hirzebruch surface as a base for the Calabi-Yau
threefold, the condition for constant coupling
corresponds to the case where the point like instantons
coalesce, giving rise to enhanced gauge group of $Sp(k)$.

\end{titlepage}
\newpage

\def\beq{\begin{equation}}
\def\eeq{\end{equation}}
\def\bea{\begin{eqnarray}}
\def\eea{\end{eqnarray}}
\renewcommand{\arraystretch}{1.5}
\def\ba{\begin{array}}
\def\ea{\end{array}}
\def\bce{\begin{center}}
\def\ece{\end{center}}
\def\nn{\noindent}
\def\nonu{\nonumber}
\def\pbx{\partial_x}

%Greek Symbols
% GREEK

\def\ptl{\partial}
\def\al{\alpha}
\def\be{\beta}
\def\ga{\gamma} 
\def\Ga{\Gamma}
\def\de{\delta} \def\De{\Delta}
\def\ep{\epsilon}
\def\vep{\varepsilon}
\def\ze{\zeta}
\def\et{\eta}
\def\th{\theta} \def\Th{\Theta}
\def\vth{\vartheta}
\def\io{\iota}
\def\ka{\kappa}
\def\la{\lambda} 
\def\La{\Lambda}
\def\rh{\rho}
\def\si{\sigma} \def\Si{\Sigma}
\def\ta{\tau}
\def\up{\upsilon} 
\def\Up{\Upsilon}
\def\ph{\phi} 
\def\Ph{\Phi}
\def\vph{\varphi}
\def\ch{\chi}
\def\ps{\psi} 
\def\Ps{\Psi}
\def\om{\omega} 
\def\Om{\Omega}

\def\lbr{\left(}
\def\rbr{\right)}
\def\half{\frac{1}{2}}

\def\vol#1{{\bf #1}}
\def\nupha#1{Nucl. Phys., \vol{#1} }
 \def\CMP#1{Comm. Math. Phys., \vol{#1} }
\def\phlta#1{Phys. Lett., \vol{#1} }
\def\phyrv#1{Phys. Rev., \vol{#1} }
\def\PRL#1{Phys. Rev. Lett., \vol{#1} }
\def\prs#1{Proc. Roc. Soc., \vol{#1} }
\def\PTP#1{Prog. Theo. Phys., \vol{#1} }
\def\SJNP#1{Sov. J. Nucl. Phys., \vol{#1} }
\def\TMP#1{Theor. Math. Phys., \vol{#1} }
\def\ANNPHY#1{Annals of Phys., \vol{#1} }
\def\PNAS#1{Proc. Natl. Acad. Sci. USA, \vol{#1} }

Our understanding of nonperturbative aspects of 
$N=2$ supersymmetric gauge theories\cite{SW} in four dimensions 
progressed very much.
Deeper understanding of the gauge dynamics comes from the study
of brane probes in string theory, where
there are also a lot of exciting developments in the conjectured 
string dualities\cite{duality}. Among these, the heterotic/type II duality has been 
studied in detail\cite{KV}. 
In fact, it has been extended to the F-theory/heterotic duality\cite{MV}
where the heterotic strings compactified on a two torus ${\bf T^2}$ is dual to 
F-theory in eight dimensions
compactified on ${\bf K3}$ which admits an elliptic fibration.
F-theory\cite{Vafa} is defined as the compactifications of type IIB string in which
the complex coupling changes over the base.
The ${\bf K3}$ surface which is a fiber space where the base is
one dimensional complex projective space ${\bf CP}^1$ and torus 
as the fiber is represented by
$y^2  = x^3 + f(z) x +g(z)$
where $z$ is the coordinate of the base ${\bf CP}^1$ and
 $f$ and $g$ are the 
polynomials of degree $8$, $12$ respectively in $z$. This describes a torus for each
point on the base ${\bf CP}^1$ labelled by the coordinate $z$.

Extension of  this to the compactification down to six dimensions
is interesting for various reasons. First of all, although the string theory
answer is rather trivial, the classical field theory on the 3-brane which
produces the answer is not so trivial and has been considered
in Ref.\cite{Aharony}.
Secondly, the orbifold limit of Calabi-Yau threefold(CY3) itself is an interesting object to study.
So far the examples which has been considered are the 
Voisin-Borcea models\cite{Vo} which are listed 
in \cite{MV} 
and are the product of a two-torus ${\bf T}^2$ and ${\bf K3}$ divided 
by a ${\bf Z}_2$ symmetry;
they correspond to type IIB compactification with a space-independent coupling constant.
It would therefore be interesting to study other examples of constant couplings
and the physics of the gauge theories arising from the string theories.

One has
the duality\cite{MV} between F-theories compactified on CY
threefold and
heterotic strings on ${\bf K3}$. 
For example, the $SO(32)$ heterotic string compactified on a 
${\bf K3}$ surface was discussed in 
Ref.\cite{small}. 
One of the interesting results is that one can obtain nonperturbative $Sp(1)$ 
extra gauge group when an instanton shrinks to zero size. 
Furthermore, when $k$ instantons collapse at the
point in the ${\bf K3}$, the $Sp(1)^{k}$ factor is replaced by the
enhanced gauge symmetry of $Sp(k)$. These results are also reproduced in 
Ref.\cite{asp2} in the context of dual theory of F-theory on an elliptically fibered
CY3. On the other hand, $E_8 \times E_8$ heterotic string compactified on
a ${\bf K3}$ has been studied in Ref.\cite{MV} where only extra massless
tensor multiplets appear as instantons shrink down to zero size. An aspect of
generic pointlike instantons for both $SO(32)$ string and 
$E_8 \times E_8$ string has been
analyzed in Refs.\cite{asp2,MV2}.
Further compactification to four dimensions
corresponds to the type II/heterotic string duality  
considered in Ref.\cite{KV}.

Sen\cite{SEN} has shown a precise relation conjectured in 
Ref.\cite{MV} between the F-theory on a 
smooth elliptic ${\bf K3}$ manifold and a type IIB orientifold on ${\bf T^2}$. 
Using the orbifold limit of ${\bf K3}$, i.e. for the simplest case of ${\bf T}^4/{\bf Z}_2$ 
where the coupling is constant over the base, new insight into
the {\bf K3} compactification was obtained. 
The points of enhanced gauge symmetries in the F-theory corresponds
to those of enhanced global $SO(8)$ symmetries in the Seiberg-Witten gauge 
theory\cite{SW}.
It has been found in Ref.\cite{DM} further that there exist other points for which
the coupling as constant i.e., $\tau=i$ or $\tau=e^{\frac{\pi i}{3}}$. At these special points,
${\bf K3}$ becomes the orbifolds of four torus
${\bf T}^4/{\bf Z}_m $ where $m=3, 4, 6$ with the base, $
{\bf T}^2/{\bf Z}_m$. At these orbifold points, a singularity analysis shows
that exceptional gauge group symmetries appear. Nontrivial superconformal
field theories for $E_{6,7,8}$ type singularities has been 
discussed\cite{mina} in the context
of Seiberg-Witten gauge theory.
Very recently,
generalizing the work of Sen\cite{SEN}, the field theory of 
3-brane probes\cite{Banks} in
a compactification of F-theory on a six torus
${\bf T}^6$ by ${\bf Z}_2 \times {\bf Z}_2$
\cite{BZ} with hodge number $(h^{11}, h^{21})=(51,3)$ 
was considered in Ref.\cite{Aharony}. 
This has an interpretation in terms of multiple 3-branes probes on an F-theory orientifold
as was discussed in Ref.\cite{DLS}.

In this paper, we do the following two things.
First, we analyze special points in the
moduli space of the compactification of the F-theory on elliptically
fibered CY3's where the coupling remains constant, along the lines
of Refs.\cite{SEN,DM}.
This is rather straightforward and can be realized as {\it other} orbifolds of six torus
${\bf T}^6/{\bf Z}_m \times {\bf Z}_n$ where $m, n=2, 3, 4, 6$.
At various types of intersection points between $G, G'=SO(8), E_{6, 7, 8}$ types of
singularities,
we find that the enhancement of gauge symmetries arises from the intersection
of two types of singularities, different from the naively expected gauge symmetry of
$G \times G'$. We find that the naive gauge symmetries get modified due to the interplay
of the singularities.
In the second part of this paper, we consider the case when the base
for the CY3 is a Hirzebruch surface $F_n$ and realize that the property of the constancy
of coupling leads to exactly the coalescence of pointlike instantons for $SO(32)$ 
heterotic string.

Let us first consider the compactification of F-theory on elliptically fibered 
CY3 where the coupling is
constant over the base. It has been found in Refs.\cite{MV,GJ2} 
that the CY3 can be described as an
elliptic fibration in the Weirstrass form 
$y^2  = x^3 + f(z,w) x +g(z,w).$
$z$ and $w$ are the coordinates on the base 
${\bf CP}^{1} \times {\bf CP}^{1} $ and $f$ and $g$ are the 
polynomials of degree $8$, $12$ respectively in each of them.
Notice that there exists an exchange symmetry when we exchange the two
${\bf CP}^{1}$'s and simultaneously the coefficient of the term, $z^{l} w^{k}$ is exchanged
with that of $z^{k} w^{l}$ in each of the terms.
The modular parameter $\tau(z,w)$ of the fiber can be written  
 in terms of the invariant $j$ function given by
\beq
j(\tau(z,w))=\frac{4(24 f(z,w))^3}{\De(z,w)},
\eeq
where the discriminant is $\De(z,w)=4(f(z,w))^3+27(g(z,w))^2$.
From now  we will  consider only the cases in which $f$ and $g$
are factorized. That is, 
$f(z,w)=\alpha f_1(z) f_2(w)$ and $g(z,w)=g_1(z) g_2(w)$ where $\alpha$
is a constant. Note that $j(\tau(z,w))$ blows up at the zeroes of
the discriminant.

The one solution for the case of
constant modulus by rescaling $y$ and $x$
and setting the overall coefficient to be 1 has been found in Ref.\cite{Aharony}.
Thus we get
\bea
&&f_1(z)=\prod_{i=1}^{4} (z-z_i)^2, \;\;\; f_2(w)=\prod_{i=1}^{4} (w-w_i)^2, \nonu \\
&&g_1(z)=\prod_{i=1}^{4} (z-z_i)^3, \;\;\; g_2(w)=\prod_{i=1}^{4} (w-w_i)^3,
\eea
where $z_i$'s and $w_i$'s are constants.
This special compactification corresponds to a configuration where
the 24 7-branes are grouped into 4 sets of 6 coincident 7-branes located at
the points, $z_1, z_2, z_3, z_4$. There exists an $SL(2,{\bf Z})$ monodromy
around each of fixed points $z_i$'s. The same is true at the points $w=w_i$
because the base is simply a product of the ${\bf CP}^1$'s.
It is obvious that we have $SO(8)$ singularities at $z=z_i$ and $w=w_i$.
The spacetime theory is an $N=1$ supersymmetric theory whose field
content was found in Refs.\cite{BZ,GM}. For example,
the open string sectors lead to $SO(8)$ gauge group for each 7-branes coming from
two ${\bf Z}_2$ factors for a total enhanced gauge symmetries $(SO(8))^4 \times
(SO(8))^4$\cite{GM}.
 
Now we continue on to carry out the same procedure
for other various subspaces of the moduli space on 
which the elliptic fiber
remains constant modulus.  As pointed out in Ref.\cite{DM}, in the limit of
$\alpha \rightarrow 0$, we get 
$j(\tau(z,w))=0$ from which $\tau(z,w)=e^{\frac{\pi i}{3}}$.
The polynomials are given by 
\bea
&& f_1(z)=0, \;\;\; g_1(z)=\prod_{i=1}^3 (z-z_i)^4, \nonu \\
&& f_2(w)=\prod_{i=1}^4 (w-w_i)^2, \;\;\; g_2(w)=\prod_{i=1}^4 (w-w_i)^3,
\label{eq:z3}
\eea
where the 12(12) zeroes of $g_1(z)(g_2(w))$
coalesce into 3(4) identical ones of order 4(2) each.
In this case, the discriminant, $\Delta(z,w)$ takes the form of
\bea
\Delta(z,w)=27 \prod_{i=1}^3 (z-z_i)^8 \prod_{j=1}^4 (w-w_j)^6.
\eea
The singularity type from 
Tate's algorithm\cite{Tate}
at a zero of the discriminant gives rise to the enhancement of gauge symmetries\cite{MV2}.
Each point $z=z_i$ on the first ${\bf CP}^1$ factor carries a deficit angle of 
$\frac{3 \pi}{2}$ all three of them together deforming the  ${\bf CP}^1$ to
${\bf T}^2/{\bf Z}_3$. For each point $w=w_i$ on the second ${\bf CP}^1$, there
is a deficit angle of $\pi$ all four of them deforming ${\bf CP}^1$ to ${\bf T}^2/{
\bf Z}_2$. This is related to orientifold of F-theory on ${\bf T}^6/{\bf Z}_3 \times {\bf Z}_2$.
In F-theory, $F_4$ gauge symmetry corresponds to the `generic' $E_6$ singularity
in the sense that the  condition on the polynomial of $g(z,w)$ splits the 
double zeroes of it in the $E_6$ gauge symmetry. Furthermore,
$G_2$ gauge symmetry corresponds to the `generic' $SO(8)$ singularity
with different constraint  on the polynomial $g(z,w)$\cite{MV2}.
Near a zero at $z_1$ the singular fiber is of $E_6$ type. 
On the other hand, $SO(8)$ type of singularity
appears near a zero at $w_1$. For simplicity, at the intersection points between these two 
`generic' singularities
the corresponding gauge group is simply  the product of $F_4 \times G_2$\cite{Ber}.
It is clear that there are no extra enhanced gauge symmetry factor because that
one blowup of the base gives rise to $II$ 
singularity\footnote{ We keep the notations of $I_n, II, III, IV, \cdots$
for the types of fiber in Kodaira's classification of singularities. (See Refs.\cite{asp3,MV})} 
on the exceptional divisor\cite{Ber},
leading to non-gauge group.
Thus the full enhanced gauge symmetry group is $(F_4)^3 \times 
(G_2)^4$ by resolving the singularity for each point of $z=z_i$ and $w=w_i$.
From an exchange symmetry between the two ${\bf CP}^1$
factors we have mentioned before we can proceed similarly for the case of compactification
F-theory on  ${\bf T}^6/{\bf Z}_2 \times {\bf Z}_3$.

When the 12 zeroes of $g_2(w)$ has coalesced into 3 identical ones of order 4
each and those of $g_1(z)$ are given as before in eq.(\ref{eq:z3}), then
we have the following:
\beq
 f_1=0, \;\;f_2=0, \;\; g_1(z)=\prod_{i=1}^3 (z-z_i)^4, \;\; g_2(w)=\prod_{i=1}^3 (w-w_i)^4,
\label{eq:33}
\eeq
where the discriminant is given by
\bea
\Delta(z,w)=27 \prod_{i=1}^3 (z-z_i)^8 \prod_{j=1}^3 (w-w_j)^8.
\eea
Of course, each point $w=w_i$ on the second ${\bf CP}^1$ factor has a deficit angle of 
$\frac{3 \pi}{2}$ all three of them together deforming the  ${\bf CP}^1$ to
${\bf T}^2/{\bf Z}_3$. This corresponds to F-theory orientifold on
${\bf T}^6/{\bf Z}_3 \times {\bf Z}_3$. Each singular fiber over the fixed point
$z_1, z_2, z_3$ and $w_1, w_2, w_3$ is
of $E_6$ type.
According to the requirement of the elliptic fibration on the blowup
surface having CY, the relation for the blown up surface restricts to the possible resolutions
and satisfies CY condition i.e. the sum of coefficients\footnote{
The coefficients $a_i$ for each type of singularity are listed in Ref.\cite{Kodaira}}  of two 
intersecting singular types in
Kodaira's list\cite{Kodaira} 
is the coefficient of singular type on the exceptional divisor plus 1\cite{Ber}.
The first blowup of the base leads to $IV$ singularity on the exceptional divisor which will
produce $SU(3)$ gauge group. The next blowups in turn appear in the intersection of
$IV \times IV^*$, known as dual for which the sum of the coefficients of $IV \times IV^*$
is always 1. So this intersection does not produce the enhancement of gauge group.
 For the intersection points of two $E_6$ singularities,
for example, at $z=z_1$ and $w=w_1$,
we get the gauge group of $E_6 \times SU(3) \times E_6$ by an extra $SU(3)$ 
factor\cite{Ber}. 
Therefore the total enhancement of  gauge symmetry group is given by
$ (E_6)^3 \times (SU(3))^3 \times (E_6)^3$. 

If the 12 zeroes of $g_2(w)$ has coalesced into 3 zeroes of order 5, 4, 3
each and those of $g_1(z)$ are the same as before like (\ref{eq:33}),  it is easy to see that
we have the following: 
\beq
f_1=0, \;\; f_2=0, \;\; g_1(z)=\prod_{i=1}^3 (z-z_i)^4, \;\; g_2(w)=(w-w_1)^5(w-w_2)^4(w-w_3)^3.
\eeq
Each point $w=w_i$ on the second ${\bf CP}^1$ factor has a deficit angle of 
$\frac{5 \pi}{3}, \frac{4 \pi}{3}$ and $\pi$ all  
together deforming the  ${\bf CP}^1$ to
${\bf T}^2/{\bf Z}_6$. We can describe this point as the
F-theory orientifold on
${\bf T}^6/{\bf Z}_3 \times {\bf Z}_6$.
In this case naive expectation is that the enhanced gauge symmetry group is the product of
$(F_4)^3 \times (E_8 \times E_6 \times G_2)$.
It is sometimes stated that this is not allowed because they violate the CY 
conditions\cite{Ber}. However, one may blow up a transversal
intersection curves of $IV^* \times II^*$ fibers {\it without} violating CY condition\cite{asp4}.
The resolutions for $IV^* \times II^*$ include $II, I_0^*, IV, I_0$. The $IV$ line cuts
the $I_0^*$ line and $II$ line. Each of these intersections induces monodromy within
the $IV$ fiber exchanging two of the three rational curves. It turns $SU(3)$ into $SU(2)$
type singularity\cite{asp4}. Therefore, $(F_4)^3 \times (G_2 \times SU(2) \times SU(3))
\times (E_8 \times E_6 \times G_2)$ gauge symmetry appears.
We can do the similar analysis for
the gauge group $ (E_8 \times E_6 \times G_2) \times (F_4)^3$ by exchanging
$z$ with $w$. 

Suppose that we go to the special point where the
each point $z=z_i$($w=w_i$) on the first(second) ${\bf CP}^1$ factor has a deficit angle of 
$\frac{5 \pi}{3}, \frac{4 \pi}{3}$ and $\pi$ all  
together deforming the  ${\bf CP}^1$ to
${\bf T}^2/{\bf Z}_6$ which indicates F-theory on ${\bf T}^6/{\bf Z}_6 \times {\bf Z}_6$.
The 12 zeroes of $g_1(z)$ and $g_2(w)$ coalesce into 3 ones
of order 5, 4, 3 each, and $f_1=f_2=0$.
% Then we get
%\bea
%&& f_1(z)=0, \;\;\; g_1(z)=(z-z_1)^5(z-z_2)^4(z-z_3)^3, \nonu \\
%&& f_2(w)=0, \;\;\; g_2(w)=(w-w_1)^5(w-w_2)^4(w-w_3)^3,
%\eea
The naive result for the enhanced gauge symmetry group is
$(E_8 \times E_6 \times G_2)^2$
which violates the CY conditions. In the case of intersection of 
$I_0^* \times II^*$  allows us to have the resolution of
$II$ and $IV$ type singularities for which the sum of the coefficients of them are less than 1. 

Let us consider the case in which
for each point $w=w_i$ on the second ${\bf CP}^1$, there
is a deficit angle of $\pi$ all four of them deforming ${\bf CP}^1$ to ${\bf T}^2/{
\bf Z}_2$, while the first ${\bf CP}^1$ factor remains unchanged.
Putting this together, we find
\bea
&& f_1(z)=0,\;\;\; g_1(z)=(z-z_1)^5(z-z_2)^4(z-z_3)^3, \nonu \\
&& f_2(w)=\prod_{i=1}^4 (w-w_i)^2, \;\;\;
g_2(w)=\prod_{i=1}^4 (w-w_i)^3,
\eea
where from the type of singularities we get 
the enhanced gauge symmetry group is
$(E_8 \times E_6 \times G_2) \times (SO(8))^4$ naively which is again
not allowed due to the CY conditions by intersecting of $II^*$ and $I_0^*$ with the similar
argument of the above.

Another possibility is as follows:
%Let us analyze the case where the 8 zeroes of $f_1(z)$ coalesce into 3 ones of order
%3, 3 and 2 and on the other hand, the 8 zeroes of $f_2(w)$ come together 4 ones of order
%2. Then it leads to
\bea
&& f_1(z)=(z-z_1)^3(z-z_2)^3(z-z_3)^2, \;\;\; g_1(z)=0, \nonu \\
&& f_2(w)=\prod_{i=1}^ 4 (w-w_i)^2, \;\;\;
g_2(w)=\prod_{i=1}^4 (w-w_i)^3,
\label{eq:z4}
\eea
which corresponds to $\tau=i$ from $j(\tau(z,w))=13824$.
This time it can be easily checked that the discriminant is given by
\bea
\Delta(z,w)=4 (z-z_1)^9(z-z_2)^9(z-z_3)^6 \prod_{i=1}^ 4 (w-w_i)^6.
\eea
The singular fiber over each fixed points $z_1, z_2$ is of $E_7$ type. The other singular
fiber over $z_3$ is of $SO(8)$ type.
At the intersection points near $z=z_1$ and $w=w_1$, the gauge group
appears to be $E_7 \times SU(2) \times SO(8)$ enhanced by an extra $SU(2)$ factor
by the fact that the first blowup for this intersection appears 
sigular type $III$ on the exceptional
divisor using the CY condition again. The intersection of $III \times III^*$ leads to a
$I_0$ type singularity which does not produce the enhanced gauge symmetry.
Hence, $(E_7 \times E_7 \times SO(7)) \times (SU(2))^2 \times (SO(8))^4$ gauge 
symmetry appears
there.

Finally, we have the case when the 8 zeroes of $f_1(z)(f_2(w))$ coalesce into 3 ones of order
3, 3 and 2, and $g_1=g_2=0$
%\bea
%&& f_1(z)=(z-z_1)^3(z-z_2)^3(z-z_3)^2, \;\;\;
%g_1(z)=0, \nonu \\
%&& f_2(w)=(w-w_1)^3(w-w_2)^3(w-w_3)^2, \;\;\; g_2(w)=0.
%\eea
For the intersection of $III^* \times III^*$, the first blow up gives rise to $I_0^*$ sigularity
on the exceptional divisor by using the CY condition as discussed above.
The arguement for the next intersection of $III^* \times I_0^*$ are given in the previous
paragraph.
Then the five resolutions correspond to $I_0, III, I_0^*, III, I_0$ in which there are
three possibilities for the type of $I_0^*$, i.e. $SO(8), SO(7)$ or $G_2$ depending on
whether the singularity is split, semi-split or non-split\cite{MV2}. 
There are three cases: no factorization in the polynomial $x^3+f(z,w)x+g(z,w)$
corresponds to non-split case,
a product of three linear factors does split case, a product of linear and quadratic factors
does semi-split case .
Only semi-split case
satisfies the anomaly factorization condition.
In this case, the gauge group appears $(E_7 \times SU(2))^2 \times SO(7)$ 
at the intersections of two $E_7$ singularities.
Then we get
 the enhanced gauge symmetry group 
$(E_7 \times E_7 \times SO(7))^2 \times(SU(2) \times SO(7) \times SU(2))^4$ 
with extra $(SU(2))^4$ factor due to the intersections
of $I_0^* \times III^*$. 

Let us note that among the various gauge groups which can appear, we have
the possibility of realizing
$(E_6)^3 \times (E_7 \times E_7 \times 
SO(7))$. For this case, the discriminant $\De(z,w)$ vanishes identically
since $f_1(z)=0$ and
$g_2(w)=0$. Thus the corresponding vacuum can not live in the F-theory moduli
space where the couplings remain constant we have studied so far but live in the full
F-theory moduli space in the sense that the coupling varies. This is also true of
the gauge group $(E_8 \times E_6 \times G_2) \times (E_7 \times E_7 \times 
SO(7))$.

We summarize our results in the following table.
\bea
\begin{array}{|c|c|}  \hline \nonu
\mbox{model} & \mbox{enhanced gauge group} \\ \hline \nonu
{\bf T}^6 /  {\bf Z}_2 \times {\bf Z}_2 & (SO(8))^4 \times (SO(8))^4 \\ \hline \nonu
{\bf T}^6 / {\bf Z}_3 \times {\bf Z}_2 & (F_4)^3 \times (G_2)^4 \\ \hline \nonu
{\bf T}^6 / {\bf Z}_3 \times {\bf Z}_3 & (E_6)^3 \times (SU(3))^3
 \times (E_6)^3 \\ \hline \nonu
{\bf T}^6 / {\bf Z}_3 \times {\bf Z}_6 & (F_4)^3 \times (G_2 \times SU(2) \times SU(3))
\times (E_8 \times E_6 \times G_2) \\ \hline \nonu
%{\bf T}^6 / {\bf Z}_6 \times {\bf Z}_6 & - \\ \hline \nonu
%{\bf T}^6 / {\bf Z}_6 \times {\bf Z}_2 & - \\ \hline \nonu
{\bf T}^6 / {\bf Z}_4 \times {\bf Z}_2 & (E_7 \times E_7 \times SO(7))
\times (SU(2))^2 \times (SO(8))^4 \\ \hline \nonu
{\bf T}^6 / {\bf Z}_4 \times {\bf Z}_4 & (E_7 \times E_7 \times SO(7))
\times (SU(2))^{12} \times (SO(7))^4 \times
 (E_7 \times E_7 \times SO(7)) \\  \hline 
\end{array}
\eea
Table 1. Possible enhancements of gauge symmetry for various F-theory orbifolds 

We can compare our findings with those in the table 8 of
Ref\cite{Ber}.
In the above table, we restricted to only the cases of intersections between $I_0^*,
II^*, III^*$ and $IV^*$ singularity types. The above five rows correspond to exactly
$I_0^* \times I_0^*, IV^* \times I_0^*, IV^* \times IV^*, 
IV^* \times II^*, III^* \times I_0^*$ and $
III^* \times III^*$ respectively when we intersect the specific two zeroes $z=z_1$ and
$w=w_1$. Our $j(\tau(z,w))$ is related to their $J$ up to constant.

In the remainder of paper, 
we would like to consider the $SO(32)$ heterotic string.
Consider the following elliptic fibration over Hirzebruch surface $F_n$
as a base for the CY3 with $z$ the
coordinate of ${\bf CP}^1$ fiber of $F_n$ and  $w$ the coordinate on the base.
This Weirstrass form may be put into the more restrictive form\cite{asp2,asp3} as follows:
\beq
y^2 =x^3+f(z,w) x +g(z,w)= 
\left( x-\beta(z,w) \right)\left( x^2 + \beta(z,w) x + \gamma (z,w)\right),
\eeq
which gives a section along $x=\beta(z,w), \ y =0$.  
The functions $\beta$ and $\gamma$ can be represented in a sufficiently
generic form as follows\cite{asp2,asp3}
\bea
\beta(z,w) = Bz^4 + Cz, \;\; \gamma(z,w) = Az^8 - 4BC z^5 -2 C^2 z^2.
\eea
Since we constrain above to be a CY space, $A$ is a polynomial of degree
$8+4n$ in $w$, $B$  is of degree $4+2n$, and $C$ is of degree $4-n$.

We consider the nonzero constant modulus case.
Using the known 
relation of the modular parameter in terms of the $j$ function, 
%\bea
%j(\tau(z,w))=\frac{4(24(-3 C^2-6 B C z^3+(A-B^2) z^6))^3}{z^{12} (A+2 B^2)^2 ((4A-
%B^2) z^6-18 B C z^3-9 C^2)},
%\eea
we get the following relation in order that $j(\tau(z,w))$ should be independent
of $z$ and $w$ for nonzero $g(z,w)$,
\beq
\frac{f^3}{g^2}=\frac{(\gamma-\beta^2)^3}{(\beta\gamma)^2}= -\frac{27}{4}.
\eeq
Notice that unlike the case of the compactification to 8 dimensions, where the
ratio of $f^3$ and $g^2$ was an arbitrary nonzero constant, here the value of the ratio
gets fixed.
For this case actually the discriminant has to vanish.
This means that we must have $A= -2B^2$ as we can see from the factorized form of 
the discriminant
\bea
\De(z,w)=4 f(z,w)^3+27 g(z,w)^2=
z^{18} (A+2 B^2)^2 ((4A-B^2) z^6-18 B C z^3-9 C^2).
\eea
For this special case, we get constant 
couplings. 
Now let us be more specific. 
Suppose $A+2B^2$ is of order $k$ in $w$. Then, we can put
\beq
A+2B^2 = \prod^{k}_{i=1} (w-w_i).
\eeq
In general the loci of the singularities $w_i$ can be different. 
The zero of the discriminant is of order $2k$ in $w$
for arbitrary value of $z$. According to Ref.\cite{Tate}, it is clear that
we have $I_{2k}$ fiber type. The gauge group related with $I_{2k}$ is $SU(2k)$.
We also observe that this case has a remarkable coincidence with the case where the
point like instantons coalesce\cite{small}, where the enhanced gauge group is $Sp(k)$.

There are other special points when $f$ or $g$ vanishes, rendering the modular
parameter of the fiber to be a constant. 
First, when $f=0$, i.e. $\beta^2 = \gamma$, we get $j(\tau(z,w))=0$,
and $\tau= e^{\frac{\pi i}{3}}$. Here we must have $C=0$ and $A=B^2$ for generic value of $z$.
Then $g= -\prod^{4+2n} _{i=1}(w-w_i)^3 z^{12}$, and if none of the $w_i$'s coincide, we get
$SO(8)$ type singularity of the discriminant and thus an enhanced gauge symmetry of 
$(G_2)^{4+2n}$ for `generic' singularity! 
Some of these might be related to the orbifold cases we discussed in the first part of the paper.
In the case of the orbifold ${\bf T}^6 /{\bf Z}_3 \times {\bf Z}_2$ as done in equation
(\ref{eq:z3}) we found that $g(z,w)=\prod_{i=1}^3 (z-z_i)^4 \prod_{j=1}^4 (w-w_j)^3$.
Merging the zeroes at $z_i$'s will produce exactly the above result for $n=0$ case.
On the other hand,
for the case of $\beta\gamma=0$ where $j(\tau(z,w))=13824$, the following relations
$B=C=0$ or $A=C=0$ hold, we get $\Delta \sim \prod^{4+2n}_{i=1} (w-w_i)^6 z^{24}$, hence
we expect again an enormous gauge enhancement of $(G_2)^{4+2n}$.
Finally when we consider the case of $A=C=0$, 
we have $f=-\prod_{i=1}^{4+2n} (w-w_i)^2 z^8$ which can be
realized as the orbifold  ${\bf T}^6 /{\bf Z}_4 \times {\bf Z}_2$ with $n=0$ 
by coalescing $z_i$'s to vanish in Eq.(\ref{eq:z4}).

Recently, the enhanced gauge group in four dimensions by $SU(n)$ singular fibers
has been studied\cite{4fold} in the context of F-theory compactifications on CY4.
Certain classes of gauge symmetry enhancement have been worked out but certainly
more works have to be done especially the cases of colliding singularities.
It would be interesting to extend our analysis for these cases.
One might also consider the compactification of F-theory on a large class of CY4's 
of the form $({\bf K3}
\times {\bf K3})/{\bf Z}_2$ in four dimensions. 
The properties of these CY4 are even less known, but for the elliptically fibered cases
with CY3 as basis will be the starting point of a future work. For example,
the orbifolds of eight torus ${\bf T}^8$ by $({\bf Z}_2)^3$
limit of this CY4s can be written as
an elliptic fibration of the form
$y^2  = x^3 + f(z,w,v) x +g(z,w,v)$
where $v, w$ and $z$ are the coordinates on $({\bf CP}^1)^3$ and $f$ and $g$ are
polynomials of degrees 8, 12 respectively in all their arguments. 
 
C.A. thanks Jaemo Park for discussions on the subject of orientifold.
We thank P.S. Aspinwall for pointing out errors in the original version of the paper.
This work is supported in part by Ministry of Education (BSRI-95-2442), 
by KOSEF (961-0201-001-2) and by CTP/SNU through the SRC program of KOSEF.


\vskip -1cm
\noindent
\begin{thebibliography}{[00]}
\baselineskip 15pt
\bibitem{SW} N. Seiberg and E. Witten, \nupha{B426} (1994) 19;
\nupha{B431} (1994) 484.
\bibitem{duality} J.H. Schwarz, hep-th/9607067; J. Polchinski, hep-th/9607050.
\bibitem{KV} S. Kachru and C. Vafa, \nupha{B450} (1995) 69.
\bibitem{MV} D.R. Morrison and C. Vafa, \nupha{B473} (1996) 74;
\nupha{B476} (1996) 437.
\bibitem{Vafa} C. Vafa, \nupha{B469} (1996) 403.
\bibitem{Aharony} O. Aharony, J. Sonnenschein, S. Yankielowicz and S. Theisen,
hep-th/9611222.
\bibitem{Vo} C. Voisin, in: Journ\'{e}es de G\'{e}om\'{e}trie Alg\'{e}brique d'Orsay(
Orsay, 1992), Ast\'{e}risque No. 218(1993), 273;
C. Borcea, in: Essays on Mirror Manifolds vol. 2, International Press (1996) Hong Kong.
\bibitem{small} E. Witten, \nupha{B460} (1996) 541.
\bibitem{asp2} P.S. Aspinwall and M. Gross, \phlta{B387} (1996) 735.
\bibitem{MV2} M. Bershadsky et al, hep-th/9605200.
\bibitem{SEN} A. Sen,  \nupha{B475} (1996) 562. 
\bibitem{DM} K. Dasgupta and S. Mukhi, \phlta{B385} (1996) 125.
\bibitem{mina} J.A. Minahan and D. Nemeschansky, hep-th/9608047;
hep-th/9610076.
\bibitem{Banks} T. Banks, M.R. Douglas and N. Seiberg, \phlta{B387} (1996) 278.
\bibitem{BZ} J.D. Blum and A. Zaffaroni, \phlta{B387} (1996) 71; A. Dabholkar and J.~Park, hep-th/9607041.
\bibitem{DLS} M.R. Douglas, D.A. Lowe and J.H. Schwarz, hep-th/9612062.
\bibitem{GJ2} E.G. Gimon and C.V. Johnson, \nupha{B479} (1996) 285.
\bibitem{GM} R. Gopakumar and S. Mukhi, hep-th/9607057.
\bibitem{Tate} J. Tate, in Modular Functions of One Variable IV, Lect. Notes in Math.
vol. 476, Springer-Verlag (1975) Berlin.
\bibitem{Ber} M. Bershadsky and A. Johansen, hep-th/9610111.
\bibitem{asp4} P.S. Aspinwall, private communications.
\bibitem{asp3} P.S. Aspinwall, hep-th/9611137.
\bibitem{Kodaira} K. Kodaira, Annals of Math., {\bf 77} (1963) 563; Annals of Math.,
{\bf 78} (1963) 1.
\bibitem{4fold} A. Klemm, B. Lian, S-S. Roan and S-T. Yau, hep-th/9701023.
\end{thebibliography}
\end{document}